\documentclass[aps,twocolumn,prl,showpacs,superscriptaddress]{revtex4}
\usepackage{graphicx} 
\usepackage{dcolumn} 
\usepackage{bm} 
\usepackage{epsfig}
\usepackage{pstcol,pst-grad}
\usepackage{amsmath}

\begin{document}

\title{Magnetization reversal in spin patterns with complex geometry}
\author{Bosiljka  Tadi\'c}
\email{Bosiljka.Tadic@ijs.si} 
\affiliation{Department for Theoretical Physics, Jozef Stefan Institute, P.O.Box 3000, SI-1001, Ljubljana, Slovenia.}
\author{Krzysztof Malarz}  
\homepage{http://home.agh.edu.pl/malarz/} 
\affiliation{Faculty of Physics and Applied Computer Science, AGH University of Science and Technology,\\al. Mickiewicza 30, PL-30059 Krak\'ow, Poland.}
\author{Krzysztof Ku{\l}akowski}
\email{kulakowski@novell.ftj.agh.edu.pl} 
\affiliation{Faculty of Physics and Applied Computer Science, AGH University of Science and Technology,\\al. Mickiewicza 30, PL-30059 Krak\'ow, Poland.}

\date{\today}

\begin{abstract}
We study field-driven dynamics of spins with antiferromagnetic interaction along the 
links of 
a complex substrate geometry, which is modeled by graphs of a 
controlled connectivity distribution. 
The magnetization reversal occurs in avalanches of spin flips, which are pinned by
the topological constraints of the underlying graph.
The hysteresis loop and avalanche sizes are analyzed and classified in terms of
graph's connectivity and clustering. 
The results are relevant for magnets with a hierarchical spatial inhomogeneity  
 and for design of nanoscale magnetic devices.
\end{abstract}


\maketitle

{\it Introduction.} Reversal processes are of great importance for
technological applications, e.g., in information processing and memory devices.
Often hysteresis curves with particular properties are required.
In this respect novel ``artificial solids'', arrays of  nanoscale magnetic particles 
\cite{nano-arrays},
quantum cellular automata \cite{QCA} and  integrated functional nanosystems  
\cite{nanowires} offer challenging possibilities, yet to be understood. 
Currently methods are being  developed for
patterning, defining and measurements of the magnetic properties at nanometer scale
\cite{nano-measurements}. Much less attention heve been devoted to the theoretical study
 of these systems \cite{theory}.

In classical disordered ferromagnets \cite{BHN_all}, 
ferroelectrics \cite{BHN_FE}
and systems with structural transformations \cite{BHN_structural} the reversal processes are 
accompanied by avalanches, which are directly related with the motion and pinning 
of the domain walls. The distribution of avalanches and the  time-series
of the magnetization bursts (generalized Barkhausen noise) exhibits universal scaling 
features, which can be used for the diagnostics of the underlying domain structure.
The pinning of the domain walls is visualized as an effect of random disorder on a
perfect crystal lattice, and commonly modeled by quenched random fields \cite{BHN_models}.
Despite of large theoretical efforts \cite{BHN_models}, the exact role of disorder 
in  the emergent universality of noise has not yet been fully understood. In general,
the stronger pinning  implies smaller avalanches and reduced coercive fields (slimmer 
hysteresis loops). 
In the exact solution \cite{DD} in the case of  Bethe lattice with large 
coordination number and weak disorder, a finite   
jump in the magnetization  persists when the system size $N\to \infty$, 
suggesting true criticality (absence of a cut-off) in the avalanche distribution 
integrated along the hysteresis loop. 

Variety of domain forms may be nucleated in 
ferroelectrics with low symmetry and long-range elastic forces \cite{Shur_new}. 
On the other hand, the patterned nanoscale magnetic structures can be assembled in 
various geometries unrestricted by a crystal symmetry \cite{nanowires}.  
The magnetostatic interactions between neighbour nanoparticles and their shape anisotropy allow
 both ferromagnetic and antiferromagnetic coupling within nanoarrays 
\cite{QCA_Nowak}. The full impact of the assembly processes and the emergent geometries 
on physical properties of the integrated nanoscale devices calls for
theoretical analysis.

The stochastic processes on complex  networks attracted much attention recently 
\cite{net_dynamics}. In general, the properties of the process depend on the complexity of the
underlying network. 
 Compared to more familiar Bravais' lattices, the network geometry makes severe 
constraints to the dynamics by restricting  the interaction pathways along the 
inhomogeneous structure of links. The network inhomogeneity proves as beneficial in the case of 
transport processes \cite{TT}.
The spin dynamics on networks has been less studied \cite{book,DS-KK}.

In this Letter we study  dynamics of spins attached to nodes of a complex network
and slowly driven by ramping of the uniform external field.
We assume {\it antiferromagnetic interactions} between spins along the links
and observe avalanches of spin flips due to topological inhomogeneity of the network.
An avalanche represents a fraction of spins that are reversed at current field value in order to 
minimize the energy within locally available geometry.
We find numerically evidence of the hysteresis loop criticality which can be explained in terms of 
the structural properties of the network.

\begin{figure}
\begin{center}
\epsfig{file=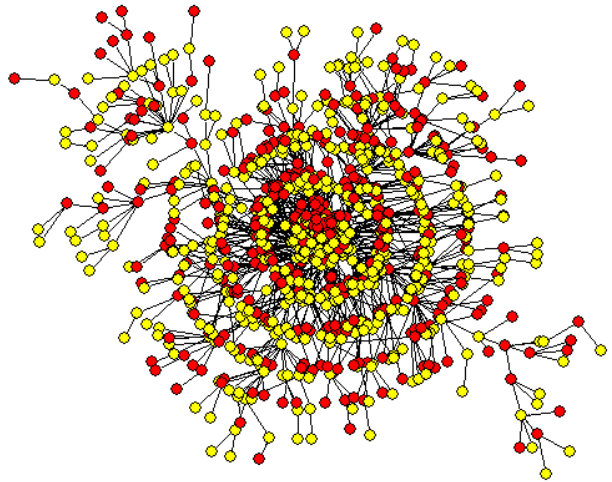, width=8.5cm}
\end{center}
\caption{\label{FIG1}
(Color online) Scale-free tree with spins attached to nodes. Different collors 
correspond to spin-up (red) and spin-down (yellow) orientation when the field 
reaches $H=0+\delta$. The snap-shot is taken during an avalanche.
} 
\end{figure}

{\it Structure.} Our model of two-state spins attached to nodes of a complex network
can be regarded as a model of an integrated nanosystem with a nontrivial architecture.
The structure is grown by systematic addition and linking of nodes \cite{book}. 
At each growth step $i$ a node is added and linked to $M$ nodes selected among $i-1$ 
preexisting nodes. Selection of a target node,  $k$, is given by a specified probability $p(k,i)$.
For $p(k,i) = 1/(i-1)$, i.e., independent of properties of target nodes, the emergent network 
is known to have an exponential degree distribution. In contrast, the preferential 
linking, where the selection probability $p(k,i)$ depends on number of already acquired links 
$q(k,i)$, are shown to lead to a scale-free degree distribution \cite{book,AB,BT}.
Here we use the probability  $p(k,i)=[1+q(k,i)/M]/2i$ to grow scale-free networks 
 for the purpose of this work.  
 We restrict our discussion to  the simplest complex networks  for the following reasons: (i) 
Their structure is well understood \cite{book,AB,BT}, in particular,  there are 
no hidden topological properties and link correlations \cite{BT};
 (ii) Varying   a single structural parameter $M$ we tune the graph clustering 
property. 
 Both scale-free network (SFN) and exponential network (EXN)  have a tree-like 
structure when  $M=1$ (an example is shown in Fig. \ref{FIG1}).
For $M>1$  cycles (disregarding the direction of links) can appear, which 
strongly depend on the linking rule.
For instance, for $M=5$, we find the average clustering coefficient \cite{AB} as 0.2803 
in SFN, and 0.0553 in EXN for network size of $N=10^3$ nodes.

\begin{figure}[htb]
\begin{center}
\epsfig{file=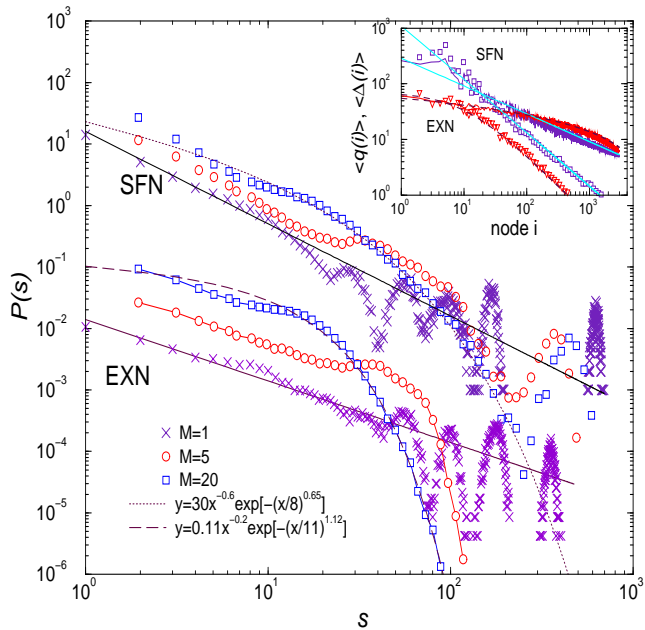, width=8.5cm, height=8.5cm}
\end{center}
\caption{\label{FIG2} Probability distribution of avalanche sizes $s$ for different clustering 
parameter $M=$1, 5, 20. Two sets of curves correspond to scale-free and exponential networks 
with  $N=10^3$ nodes, averaged over $N_s =10^3$ network realizations.
Inset: Profile of the local connectivity $\langle q(i)\rangle$ (full lines) and number of triangles 
$\Delta(i)$ (symbols) for the two network types,  $N=3000$, $N_s=10$. 
Fits: dashed lines $\langle q(i)\rangle \propto \log (10^4/i)$, $\langle\Delta (i)\rangle =
70(1+i/10)^{-1.12}$;  
solid lines slopes: $-0.5$ and $-1$. 
}
\end{figure}

The network sparseness and structural inhomogeneity are the feature which can affect 
the spin dynamics.  In the two network types introduced above, the profiles of the local 
connectivity (number of links attached to a node) are shown in the inset to Fig. \ref{FIG2}, 
where $i$ denotes the order of addition of the node to the network. In the case of SFN of
 size $N$ the profile is given by  the power-law dependence $\langle q(i,N)\rangle \propto 
(N/i)^{\gamma}$, with $\gamma=1/2$  related to the emergent power-law degree distribution 
 $P(q)\propto q^{-3}$ 
( see \cite{book,BT} for more general scaling relations). In the case of EXN we have  
$\langle q(i,N)\rangle \propto \log (N/i)$. When $M>1$ the number of the elementary triangles 
or {\it clustering} $\langle \Delta(i)\rangle$  at a node $i$ (i.e., number of direct links 
among nearst-neighbors of $i$) also varies through the network. The clustering  profiles 
for $M=5$ in  SFN and EXN, 
exhibiting power-law tails, are also shown in the inset to Fig. \ref{FIG2}.


{\it Spin Dynamics.} A spin, $S=\pm 1$, is associated with each node
on the network.
 The spins interact along the links connecting neighbouring nodes.
 We assume uniform anti-ferromagnetic interaction $J=-1$ and  uniform external field $H$
according to the Hamiltonian 
\begin{equation}
{\cal H} \equiv -\sum_i h_i S_i = -\sum_{i,j>i} J C_{ij} S_i S_j - H\sum_i S_i \ .
\label{Ham}
\end{equation}
The sum over pairs of nodes is restricted to the locally available links, which 
are given by the positive elements $C_{ij}$ of the adjacency matrix $\mathbf{C}$ of the graph. 
The interaction along these links is symmetrical.
Therefore, the local field $h_i$, which is defined in Eq. \eqref{Ham}, varies with the local 
connectivity profile $\langle q(i,N)\rangle$ of the graph.

\begin{figure*}[bht]
\begin{center}
\hskip 1.8 true cm
\epsfig{file=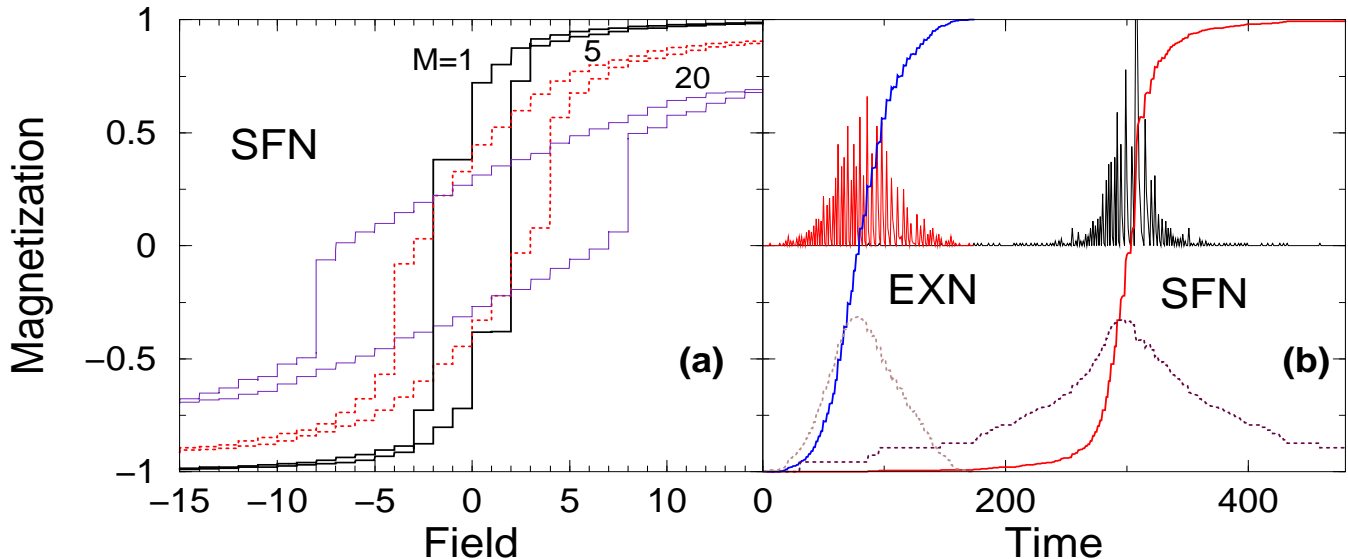, width=15.5cm, height=6cm}
\end{center}
\vskip 0.5 true cm
\caption{\label{FIG3}
(a) Hysteresis loop for the scale-free network of different clustering parameter $M=1$ (tree), 
5, and 20.
(b) Time dependence of the magnetization, number of flips (Barkhausen noise), and 
domain-wall density (broken lines, shifted by $-1$ for better view), for two types of 
network structures  and for $M=5$.}
\end{figure*}

We start from a negative saturation state in the field $H = -H_{\text{max}}$,
where $H_{\text{max}} = q_{\text{max}} + \delta$ is determined  by the
largest connectivity $q_{\text{max}}$ (at hub node) for each network realization, 
and $\delta$ is a small shift which prevents $h_i=0$.
Slowly increasing  external field is outbalanced at some node where the interaction with other spins is the strongest.
Flipping the spin at that node may launch a cascade of flips along the connections of the graph.
The avalanche of flips is stopped when the energy is minimized at current field value.
Then the field is increased 
again. This produces the step-wise increase of the magnetization with time and with field,
as shown in Fig. \ref{FIG3}, in a manner known as Barkhausen effect in systems with domain 
structure.
Compared to the classical Barkhausen avalanches 
in driven disordered systems \cite{BHN_all,BHN_FE,BHN_structural,BHN_models},  apart 
from the {\it absence of disorder}, the following important differences will be 
pointed out: (i) Only integer values of the local fields $h_i$ occur, making the 
lowest driving rate $\Delta H =1$ finite \cite{comment_rate}; 
(ii) With zero temperature dynamics that we use, back flips are possible due to the
antiferromagnetic coupling; (iii) Disconnected avalanches may appear. We consider 
number of flips before 
a cascade stops as a measure of the avalanche size $s$. 
The net magnetization changes may differ from the number of flips
during one driving event. We  define a time step as one update of the whole network.
In the presence of the antiferromagnetic interactions the sequential updates  
are suitable for all  $M$ values. The distributions of avalanche 
sizes recorded along the ascending branch of hysteresis for the two network 
topologies and different clustering parameter $M$ are shown 
in the main Fig.\ \ref{FIG2}.

{\it Hysteresis-Loop Criticality.}
We first discuss in detail the case $M=1$, corresponding to a tree-like structures. 
As displayed in Fig.\ \ref{FIG2}, the distributions 
of avalanche sizes exhibit a typical oscillatory behavior superimposed on an underlying
 power-law dependences, with different slopes in SFN and EXN.
The characteristic peaks correspond to  collective spin flips \cite{KM_etal} on a given  
depth layer of the tree (cf. Fig.\ref{FIG1}). In both 
cases the depth of a graph is proportional to $D\propto \log (N)$
(with different proportionality constants), however, the  layers are
more densely populated in the EXN than in the SFN. For instance, the number of 
spins which are directly linked to the spin at the hub node (first layer) in SFN  is 
sixfold larger than in the EXN of the same size (cf. inset to Fig. \ref{FIG2}). The 
population differences  along layers  roughly correspond to 
the  relative positions of peaks in Fig. \ref{FIG2}.

Typically, spin at the hub  is reversed early, then the structure around the hub 
remains stable until the field reaches very high values.
This is compatible with the time picture, (Fig. \ref{FIG3}(b) shows $M=5$ case), where the 
magnetization increase in the SFN takes much longer time, compared with  the EXN. 
A snap shot of the spin-up and spin-down populations corresponding to the middle of the hysteresis 
loop in SFN is displayed in Fig. \ref{FIG1}.
Broad network connectivity distribution does not allow definition of magnetic sublattices, such 
as familiar on Bravais' crystal structures. Moreover, 
the evolution of the {\it density of domain walls}
has a characteristic profile for each network structure, as shown in Fig. \ref{FIG3}(b).
This results in the hysteresis loop of particular properties. 
 Fig. \ref{FIG3}(a) show the hysteresis loop for the SFN, which is dominated by 
the presence of the hub: Both, the extended narrow tails, 
and the discontinuity jump at the critical field, are attributed to the large connectivity of 
the hub node.

With increased clustering of networks, i.e., when the parameter $M>1$,  closed loops  of 
interacting spins appear. The nearest-neighbor triangles have largest impact to the spin dynamics, 
being incompatible with the antiferromagnetic interactions. This leads to {\it frustration} of 
spins, a familiar concept in the theory of spin-glasses \cite{SG_frustrations}. 
However, the observed inhomogeneous distribution of
 triangles on the scale-free graph, as shown in the inset 
to Fig.\ \ref{FIG2}, affects the hysteresis loop 
criticality in different manner, compared with the case of spin-glasses 
\cite{comment_weissman}. In particular, the critical field $H_c$ of the magnetization 
reversal
{\it increases} with the clustering on the scale-free network and a finite 
magnetization discontinuity  persists  even for very large clustering (see Fig.\ \ref{FIG3}(a)). 
On the other hand, the absence of the hub nodes in a mild logarithmic profile of the EXN,
results in a narrow hysteresis loop and the absence of the large discontinuity for all
 $M>1$, more similar to the case of Bravais' solids \cite{comment_weissman}. 

Increased clustering results in stronger pinning \cite{DS-KK}. 
Consequently, the cut-off decreases in the avalanche size distribution, with
gradual smearing of the oscillations. The distributions approach the 
 form that can be  fitted by a stretched exponential function for $M=20$. 
The qualitatively same behavior with different parameters is found in both types of network 
structures (cf. Fig. \ref{FIG2}). 
 The distributions of avalanche durations are broad with increased cut-offs 
 for large clustering.
However, owing to small depth  (``small-world'' feature) of the networks, the overall
range of the distributions is short (about one decade) to determine a precise functional 
form \cite{KM_etal}.

In order to understand the nature of the hysteresis loop criticality in driven disordered
ferromagnets, recently several exact results on Bethe lattice \cite{DD} and other 
sophisticated theoretical approaches were applied \cite{DD_bpercol}. 
 It is likely that these approaches, based on the 
analogy of the Barkhausen avalanches with the stochastic branching processes and 
bootstrap percolation, can be useful for the study of the hysteresis curves on 
networks, with a necessary adjustments that will take into account  
two essentially different points: 
(i) antiferromagnetic interactions and (ii) power-law connectivity profile. 
In particular, based on our numerical results in Fig. \ref{FIG2}, we expect that 
true criticality can  be proved in the case of tree structures ($M=1$) with the scaling 
exponents of the 
avalanche sizes expressible in terms of the connectivity profile $\gamma$ as 
$\tau_s = 1 + \gamma  =1.5$ 
for the scale-free, and $\tau_s =1$ for the logarithmic profile in the exponential tree.


{\it Conclusions.}
We have shown that topological constraints due to 
complex  geometry of network  and  antiferromagnetically coupled spins leads to
the avalanche-like magnetization reversal and hysteresis-loop criticality {\it without
disorder}. The broad distribution of the network connectivity and the presence of the hubs
in the scale-free structures
 leads to unexpected features of the magnetization reversal:
 increased clustering promotes stronger pinning and shorter 
avalanches, but at the same time increases the critical field and preserves a finite
discontinuity in the magnetization.  
These features may require renewed theoretical concepts. In practical terms, 
our results suggest that the control of the pattern geometry implies the control of 
the hysteresis curve properties within  an {\it enlarged range  of parameters}. 
Our simplified model with  ``free graphs'' may initiate study of more realistic
 models and practical analysis of integrated  nanosystems   
with rich structural and magnetic characteristics.

B.T.  acknowledges  support by project  No.   P1-0044  of  the  Ministry  of
Education,  Science   and  Sports  (Slovenia). K.M. thanks hospitality at J. 
Stefan Institute, Ljubljana.


\end{document}